\documentclass[graybox]{svmult}

\usepackage{type1cm}        
\usepackage{makeidx}         
\usepackage{graphicx}        
\usepackage{multicol}        
\usepackage[bottom]{footmisc}

\usepackage{newtxtext}       %
\usepackage{newtxmath}       

\makeindex             

\begin{document}

\title*{Physical conditions and chemical abundances in photoionized nebulae from optical spectra}
\titlerunning{Physical conditions and chemical abundances in photoionized nebulae}
\author{Jorge Garc\'{\i}a-Rojas}
\institute{Jorge Garc\'{\i}a-Rojas \at Instituto de Astrof\'{\i}sica de Canarias, E-38200, La Laguna, Tenerife, Spain. Universidad de La Laguna. Depart. de Astrof\'{\i}sica, E-38206, La Laguna, Tenerife, Spain \email{jogarcia@iac.es}}
%
%
\maketitle

\abstract{This chapter presents a review on the latest advances in the computation of physical conditions and chemical abundances of elements present in photoionized gas (H~{\sc ii} regions and planetary nebulae). The arrival of highly sensitive spectrographs attached to large telescopes and the development of more sophisticated and detailed atomic data calculations and ionization correction factors have helped to raise the number of ionic species studied in photoionized nebulae in the last years, as well as to reduce the uncertainties in the computed abundances. Special attention will be given to the detection of very faint lines such as heavy-element recombination lines of C, N and O in H~{\sc ii} regions and planetary nebulae, and collisionally excited lines of neutron-capture elements (Z$>$30) in planetary nebulae. \newline\indent}

\section{A very brief introduction on emission line spectra of photoionized nebulae}
\label{sec:intro}

Photoionized nebulae (i. e. H~{\sc ii} regions and planetary nebulae) are among the most ``photogenic'' objects in the sky. Given their relatively high surface-brightness they are easily accessible, even for non-professional telescopes. This allowed earliest visual spectroscopic observations by William Huggins and  William A. Miller (\cite{hugginsmiller64}) who obtained the first spectrum of a planetary nebula (The Cat's Eye Nebula), where they detected a bright emission line coming from a mysterious element that Margaret L. Huggins (\cite{huggins98}) called ``nebulium''. Several decades later, Ira S. Bowen (\cite{bowen27}) showed that this emission was produced by doubly ionized oxygen (O$^{2+}$) at extremely low densities. An historical review on the early steps of the study of the physics of gaseous nebulae was provided by Donald E. Osterbrock (\cite{osterbrock88}) who used a seminal paper by Bowen (\cite{bowen27b}) as the starting point for a review of nebular astrophysics.

Photoionized nebulae are excited by the strong ultraviolet (UV) radiation of hot stars ($T_{eff}\geq 25-30 kK$) which produce photons with energy that could be above the ionization threshold of the gas particles and hence, ionize them releasing a free electron. The probability of occurrence of this phenomenon depends on the photoionization cross-section which, in turn, depends on the energy of the photon and the target being considered. Once ionized, the gas particles tend to recombine with the free electrons, and eventually an equilibrium stage is established in which the rate of ionization equals the rate of recombination for each species (see \cite{osterbrockferland06}). 
 
\begin{figure}  
\sidecaption
\hspace{0.25cm}
\includegraphics[width=\textwidth]{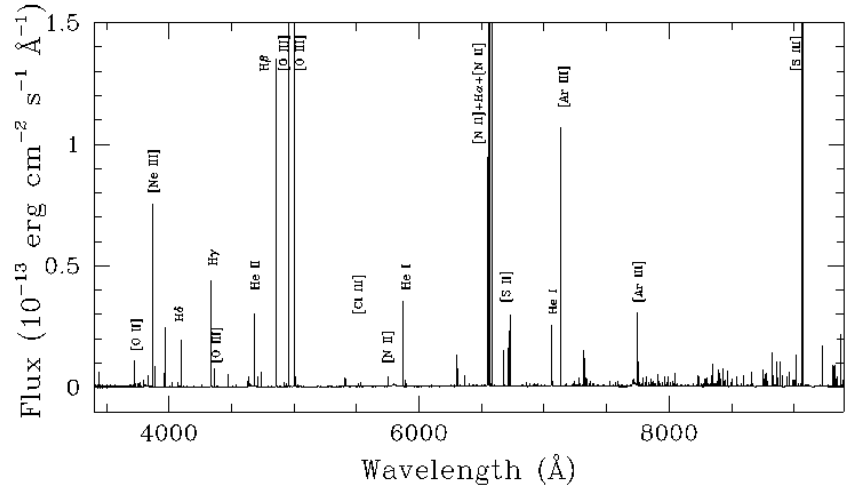}
\caption{Section of the very deep spectrum of the PN Hb\,4 analysed in \cite{garciarojasetal12, garciarojasetal13} showing bright RLs of H~{\sc i} and He~{\sc i} and CELs of different ions of O, N, S and Ne. }
\label{fig:1}
\end{figure}

The optical spectra of photoionized nebulae are dominated by emission lines, which are formed when atoms or ions make a transition from one bound electronic state to another bound state at a lower energy via spontaneous emission. These bound electrons can be excited either by free electrons colliding with the atom/ion, or by absorption of a photon. However, the background radiation field in the interstellar medium in generally not strong enough for excitation by photon absorption to be significant (see chapter 5 of \cite{kwok07}) and therefore, the only way of having a bound electron in an excited state is by collisional excitation from a lower state, which subsequent radiative decays to lower levels originating the collisionally excited lines (hereinafter CELs), or owing to a recombination between a free electron and an ion, which is the mechanism behind the emission of recombination lines (hereinafter RLs). Given that the abundance of H and He ions are several orders of magnitude higher than that of heavier elements, one can instinctively assume that the emission spectra will be dominated by H and He lines, which is not the case. In photoionized nebulae the peak of the energy distribution of free electrons is of the order of 1 eV. Ions of heavy atoms like N, O, Ne, S, Cl, Ar, etc. have electronic structures with low-lying electronic states in the range of fractions to few eV from the ground state and can therefore be effectively excited by collisions. On the other hand, for H and He ions, the gap between the ground state and the first excited state is very large and cannot be excited by collisions, but by recombination.  Fig.~\ref{fig:1} shows a typical optical spectrum of a photoionized nebulae (in this case the planetary nebula Hb\,4); remarkably bright H and He RLs and CELs of different ionic species of N, O, Ne, S, Cl and Ar are labelled. 

\begin{figure}  
\sidecaption
\hspace{0.25cm}
\includegraphics[width=\textwidth]{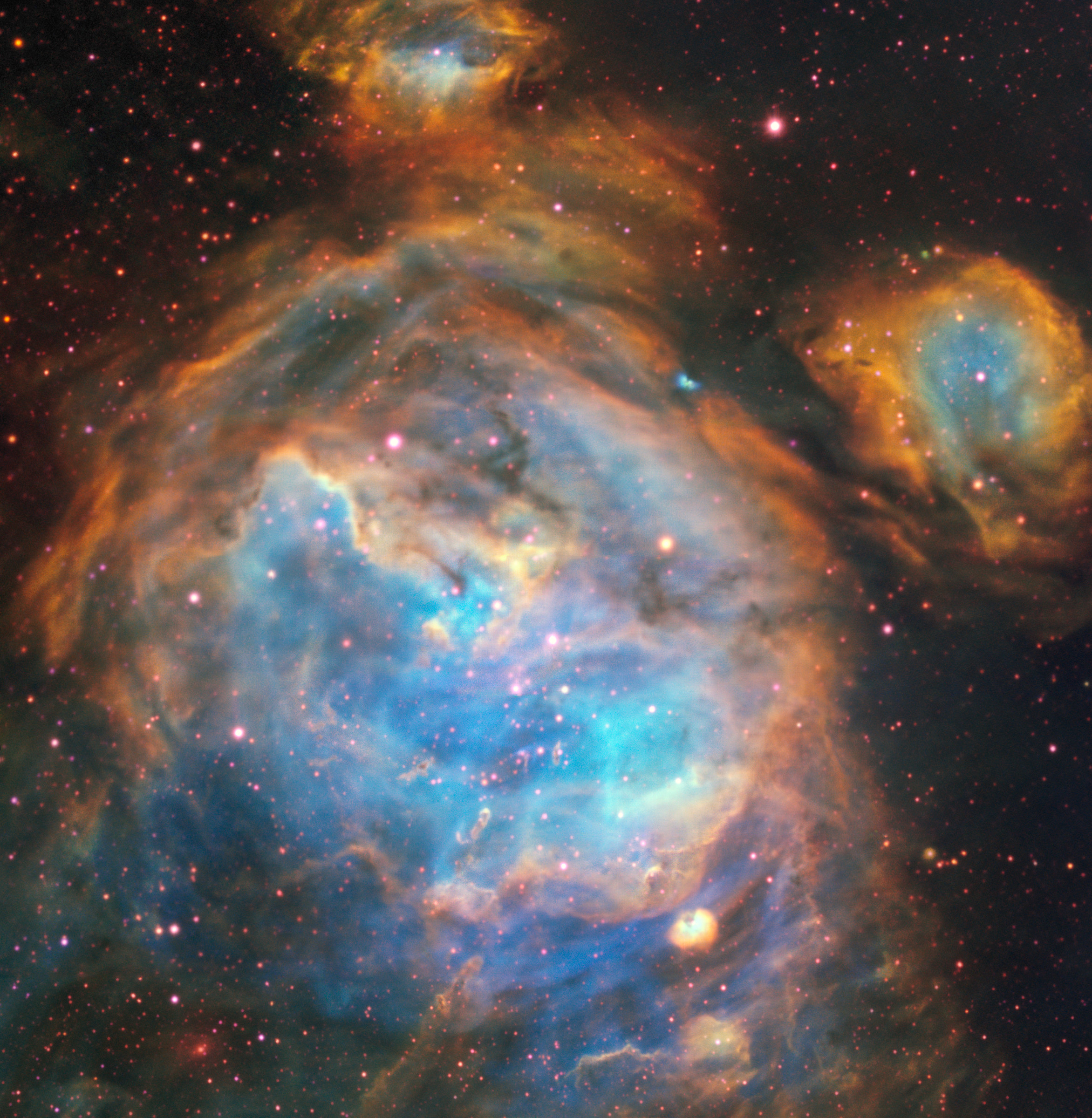}
\caption{``Bubbles of Brand New Stars'' Composite image of a star-forming region in the Large Magellanic Cloud (LMC) captured by the Multi Unit Spectroscopic Explorer (MUSE) instrument on ESO's Very Large Telescope (VLT). The following colour code was used: [O~{\sc iii}] $\lambda$5007 (blue).H$\alpha$ (yellow), [S~{\sc ii}] $\lambda$6731 (red). The field-of-view of the image is 7.82 $\times$ 8.00 arcminutes$^2$. North is 180.2$\deg$ left of vertical. Credit: ESO, A. McLeod et al.}
\label{fig:2}
\end{figure}

Therefore, the  spectrum of a photoionized nebula is dominated by the emission of RLs of H and He (the most abundant elements) and CELs of heavier elements. The combination of narrow-band images taken in the brightest emission lines allows to construct the beautiful coloured images of photoionized nebulae (see Fig.~\ref{fig:2}) from which we can have a first sketch of the ionization structure of the photoionized region. In Fig.~\ref{fig:2} we show a 3 narrow-band filter combined image of a star-forming region in the Large Magellanic Cloud where is clear that the emission of [O~{\sc iii}] is more internally located than the emission from [S~{\sc ii}]. 

Although the emission line spectra from H~{\sc ii} regions and planetary nebulae (hereinafter, PNe) are roughly similar, there are some remarkable differences between them. H~{\sc ii} regions are large (tens of parsecs), massive (generally between $10^2-10^3$ M$_{\odot}$) regions of gas that are ionized by the ultraviolet (UV) radiation emitted by recently formed OB-type massive stars with typical effective temperatures between 25-50 kK; in general, these stars are not hot enough to ionize nebular He~{\sc ii}, whose ionization potential is $h\nu=54.4$ eV. However, there are exceptions to this rule, especially in the integrated spectra of blue compact dwarf galaxies (BCDs), Wolf-Rayet (WR) galaxies and a couple of nebulae in the Local Group, associated to WR stars. On the other hand, PNe are much smaller ($10^{-1}$ pc) and less massive ($\sim$10$^{-1}$ M$_{\odot}$) nebulae that are excited by central stars which are generally hotter (central stars can reach temperatures as high as 250 kK); therefore, there will be ionizing photons with enough energy to ionize high-excitation species and hence, producing qualitatively different spectra than that of H~{\sc ii} regions, showing  emission lines of He~{\sc ii}, [Ne~{\sc v}], [Ar~{\sc v}], [Fe~{\sc v}], and even more excited species. 

\subsection{Why are abundances in photoionized nebulae important in astrophysics?}
\label{sec:importance}

The analysis of emission line spectra of photoionized nebulae allows us to determine the chemical composition of the interstellar medium (ISM) from the solar neighbourhood to the high redshift
star-forming galaxies. It stands as an essential tool for our knowledge of stellar nucleosynthesis and the cosmic chemical evolution. Since the early achievements in spectrophotometry of photoionized nebulae, the quality of deep optical and near-infrared spectrophotometric data of PNe has increased significantly mainly thanks to both the development of more efficient instruments and to the advent of large aperture (8m-10m-class) ground-based telescopes. In this sense, the future installation of giant-class ones (diameters 30-50m) opens new horizons in the field of nebular spectroscopy. The detection of very faint emission lines in ionized nebulae as auroral CELs in faint, distant or high-metallicity objects; optical recombination lines (hereinafter, ORLs) of heavy-element ions or CELs of trans-iron neutron-capture elements are becoming a routine fact and provide new information of paramount interest in many different areas of astrophysics.

H~{\sc ii} regions can be observed at considerable distances in the Universe and hence, are crucial to determine the chemical composition of the interestellar medium (ISM) in the extragalactic domain. Since H~{\sc ii} regions lie where star formation is occurring, chemical abundances computed in H~{\sc ii} regions are probes to trace the present-day chemical composition of the ISM. In particular, the study of radial variations of chemical abundances along galactic discs in spiral galaxies are essential observational constraints for chemical evolution models, and precise determinations of chemical abundances in low-metallicity dwarf galaxies, can permit to determine the primordial abundance of helium owing to Big Bang nucleosynthesis  (see \cite{peimbertetal17} and references therein). The global picture of abundances in PNe is more complicated because for elements that are supposed to be not modified, such as O and $\alpha$-elements, the computed abundances reflect the chemical conditions in the cloud where the progenitor star was formed, while the chemical abundances of N, C, or neutron-capture elements, that could be modified during the cycle of life of low-to-intermediate mass stars allow us to constrain the nucleosynthetic processes in these stars.

\subsection{Recent reviews on chemical abundance determinations}
\label{sec:reviews}

Recently, two tutorials focused on the determination of ionized gaseous nebulae abundances have been released (\cite{peimbertetal17} and \cite{perezmontero17}) although with different points of view. In the former, \cite{peimbertetal17}  give a brief review on the physics basics of abundance determinations, like local ionization and local thermal equilibrium, emission line mechanisms and on the calculation of physical conditions and ionic and elemental abundance determinations from observations; these authors also review recent results in abundance determinations in both H~{\sc ii} regions and PNe. However the review is quite focused to the abundance discrepancy problem from the point of view of temperature fluctuations (see Section~\ref{sec:adf}). In \cite{perezmontero17}, the focus is on the determination of abundances in extragalactic H~{\sc ii} regions from the direct method (when electron temperature, $T_e$, and electron density, $n_e$, diagnostic lines are available) and in the use of some strong-line methods calibrated using the direct method (see Section~\ref{sec:strong}). 

Further comprehensive tutorials on abundance determinations are those by \cite{stasinska02} and \cite{stasinska07} where the theoretical background of photoionized nebulae is treated in more detail, and particular emphasis is given to the description of  line formation mechanisms, transfer of radiation, as well as to the use of empirical diagnostics based on emission lines and determination of chemical abundances using photoionization models. It is not the scope of this chapter to repeat the basic concepts of the physics of photoionized nebulae, which have been described in different detail in the aforementioned tutorials. Moreover, for a much more detailed description of such processes, we refer the reader to the canonical book of photoionized nebulae: ``Astrophysics of gaseous nebulae and active Galactic nuclei'' (\cite{osterbrockferland06}).

In the following sections I will focus on recent advances in chemical abundances determinations in photoionized nebulae from the analysis of deep optical and near-infrared spectra, from an observational point of view. Due to space limitations, I refer the reader to \cite{stasinska02, stasinska07} for an overview on abundance determinations based on photoionization model fitting. Similarly, the strong line methods to determine abundances in the extragalactic domain (from giant H~{\sc ii} regions to high-redshift galaxies) will be only briefly discussed in Section~\ref{sec:strong}.  

\section{Observational spectroscopic data: the first step to obtain reliable abundances.}

\begin{figure}  
\sidecaption
\hspace{0.25cm}
\includegraphics[width=\textwidth]{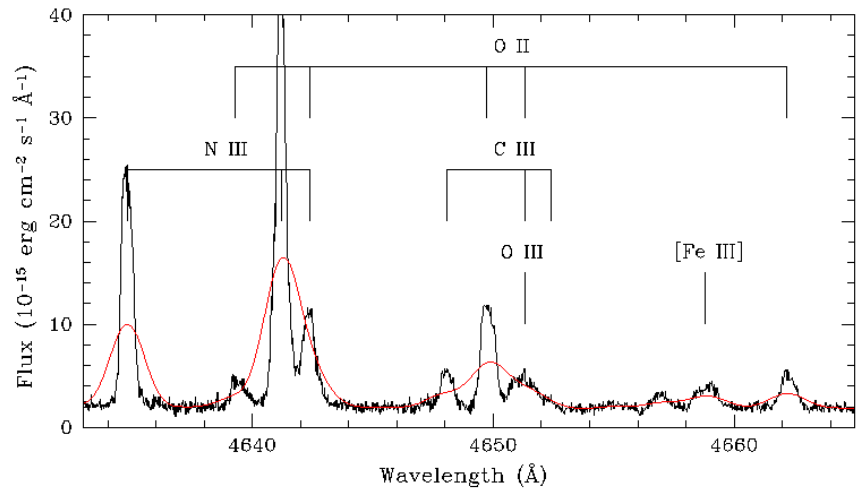}
\caption{Portion of a high-resolution (R$\sim$15000) spectra of the high-excitation PN H\,1-50 showing the zone where the multiplet 1 O~{\sc ii} lines lie. Overplotted in red is the same spectra degraded to a resolution of R$\sim$3500. As it can be shown the high-resolution of the original spectra allows one to deblend several very close permitted emission lines of C, O, and N that would have remained hidden in the low-resolution spectra. Data originally published in \cite{garciarojasetal18}.}
\label{fig:3}
\end{figure}

In the last years, the number of deep high-quality spectra of photoionized nebulae has increased significantly, allowing the detection of very faint emission lines (see e.~g. \cite{liuetal00, tsamisetal03, tsamisetal03b, estebanetal04, garciarojasetal04, wessonetal05, fangliu11, garciarojasetal12, garciarojasetal15, madonnaetal17, wessonetal18}) and the computation of, in principle, very reliable chemical abundances. The advantages of obtaining deep and high resolution spectra of photoionized regions are clear because one can easily isolate faint lines that in lower resolution spectra would be blended and go unnoticed.   As an illustration, in Fig.~\ref{fig:3} we show an excerpt of the spectrum of the high-excitation PN H\,1-50 analysed in \cite{garciarojasetal18} with the same spectrum downgraded to a lower resolution overplotted in red. Several permitted lines of O, N, and C would have remained hidden in the low-resolution spectra and \emph{ad-hoc} atomic physics would have been needed to estimate their fluxes. In the last years, several groups have provided a large sample of deep, high-resolution spectra of both Galactic and extragalactic H~{\sc ii} regions and PNe (see e.~g. \cite{toribiosanciprianoetal16, toribiosanciprianoetal17, garciarojasetal18} and the compilation made by \cite{mcnabbetal13}).

\begin{figure}  
\sidecaption
\hspace{0.25cm}
\includegraphics[width=\textwidth]{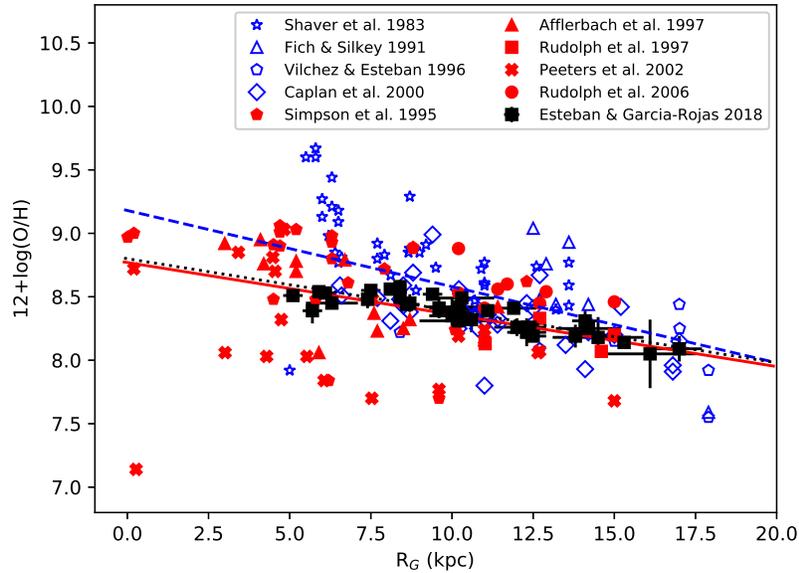}
\caption{Radial oxygen abundance gradient in the MW using abundances derived from optical lines (blue points) or from far-IR lines (red points) compiled by \cite{rudolphetal06}. The data from \cite{estebangarciarojas18} obtained through deep spectra taken in 8m-10m class telescopes are overplotted (black squares). The fits for each set of data are represented by lines with the same colour than the data points. References of the original data compiled by \cite{rudolphetal06} are shown in the legend.}
\label{fig:4}
\end{figure}

However, deep, high signal-to-noise, high-resolution spectra are not the panacea. \cite{rodriguez19} has recently shown that the effects of observational uncertainties can be very important even making use of high quality spectra, owing to the high number of sources of uncertainty that are acting in the process, which include: assumptions in the nebular structure, atomic data (see section~\ref{sec:atomic}), atmospheric differential refraction, telluric absorption and emission, flux calibration, extinction correction, blends with unknown lines, etc. Therefore, a careful data reduction procedure should be carried out to obtain reliable results. Additionally, an homogeneous analysis determining physical conditions and chemical abundances from the same set of spectra is mandatory if one want to compute precise abundances. For instance, many studies devoted to study the radial abundance gradients have made use of  physical conditions derived from radio recombination lines combined with and optical or infrared lines to compute abundances; these approach has been used in several seminal papers on the Galactic abundance gradient (see section~\ref{sec:grad}; however, it can introduce systematic uncertainties owing to the different areas of the nebula covered in the different wavelength ranges. Additionally, we should use a set of appropriate lines to compute the abundances; as an example, computing O$^+$/H$^+$ ratios from the trans-auroral [O~{\sc ii}] $\lambda\lambda$7320+30 lines could introduce undesired uncertainties because these lines could be strongly affected by telluric emission, and are also very sensitive to electron density and temperature. To illustrate these effects, in Fig.~\ref{fig:4} we show an adaptation of Fig.~5 of \cite{rudolphetal06} where the radial oxygen abundance gradient making a consistent analysis of several data sets is presented. Physical conditions have been derived from both radio and optical diagnostics, and abundances have been derived using optical CELs of oxygen (blue points) or far-IR fine-structure CELs of oxygen (red points). As can be seen, both data sets show significant scatter in the oxygen abundance at a given Galactocentric distance, which can be interpreted as an ``intrinsic scatter'' owing to the gas not being well mixed (\cite{afflerbachetal97, rudolphetal06, rosolowskisimon08}). However, high-quality observations seem to rule out this interpretation. In Fig.~\ref{fig:4} the abundances computed by \cite{estebangarciarojas18} from an homogeneous analysis of optical spectrophotometric data of 35 H~{\sc ii} regions with direct determinations of the electron temperature have been overplotted on the \cite{rudolphetal06} sample. As it is clearly shown, the scatter in the oxygen abundance is reduced significantly and is not substantially larger than the observational uncertainties, indicating that oxygen seems to be well mixed in the ISM at a given distance along the Galactic disc. Moreover, \cite{bresolin11} showed from the analysis of  high-quality spectra with high signal-to-noise auroral [O~{\sc iii}] line detections in H~{\sc ii} regions in the inner parts of M\,33, a much lower scatter than that found by \cite{rosolowskisimon08}; this author also found no evidence for significant azimuthal variations in the H~{\sc ii} region metallicity distributions, ruling out large anomalies in the mixing of the gas.

Finally, one has to take into account some biases that the direct method can have. \cite{stasinska05} discussed about the limitations of the direct method to determine O abundances in giant H~{\sc ii} regions at metallicities larger than solar. This author used {\it ab-initio} photoionization models of giant H~{\sc ii} regions, and applied to the models the same methods as used for real objects to test the direct method. The global result of this study was that for log(O/H)+12 larger than 8.7 (i.e. larger than the solar value), the computed O/H values were below the ones implied by the photoionization models owing to strong temperature gradients present in giant H~{\sc ii} regions. Finally, \cite{stasinska05} propose that PNe, which are not affected by these biases, could be potential probes of the metallicity of the interstellar medium in the internal parts of spiral galaxies as well as in metal-rich elliptical galaxies. However, in Section~\ref{sec:oxygen} we will discuss that this idea should be taken with some caution. 

\section{Determination of physical conditions and ionic abundances}
\label{sec:phys_cond}

\subsection{The direct method}
\label{sec:direct}

The most popular way to compute the chemical abundances of the elements that are present in a photoionized gas is the so-called direct method. This method makes use of CELs intensities of different ionic species of elements like N, O, S, Ne, Cl, Ar, Ne, Fe, etc., and involves the determination of the physical conditions (temperature and electron density) in the emitting plasma. In the conditions prevailing in photoionized nebulae like H~{\sc ii} regions and PNe, most of the observed emission lines are optically thin\footnote{An emission line is said to be optically thick if on average a photon emitted cannot pass through the ISM without absorption. Conversely, an emission line is said to be optically thin if we can see the radiation coming from behind the nebula (i.~e. it is not absorbed).} with the exception of some resonance UV lines and some fine-structure IR lines (see \cite{stasinska02}) making their use for abundance determinations very robust. 

In the analysis of photoionized nebulae it is usually assumed that the physical conditions are homogeneous in the photoionized region. Under these assumptions one can compute electron temperature and density by using sensitive line ratios. Electron temperature ($T_e$) and density ($n_e$) in nebulae are represented by the kinetic energy and density of the free electrons in the photoionized gas. Some CEL intensity ratios of ions of common elements like O, N, S or Ar depend on the physical conditions of the gas, and are useful to calculate $T_e$ and $n_e$ (see Section 3.5 of \cite{peimbertetal17} for more details). In particular, the intensity ratios of emission lines of a given ion that originates in very different energy levels, are sensitive to $T_e$ and almost independent on $n_e$, since the populations of the different atomic levels are strongly dependent on the kinetic energy of the colliding free electrons. Typical optical electron temperature diagnostics are: [N~{\sc ii}] $\lambda$5754/$\lambda$6583, [O~{\sc ii}] $\lambda\lambda$7320+30/$\lambda\lambda$3726+29, [O~{\sc iii}] $\lambda$4363/$\lambda$5007,  [Ar~{\sc iii}] $\lambda$5191/$\lambda$7531 or [S~{\sc iii}] $\lambda$6312/$\lambda$9531. Therefore, determination of chemical abundances making use of the direct method in optical spectra requires the detection of faint auroral lines, which correspond to transitions from the state $^1$S to $^1$D and are very $T_e$-sensitive. The detection of such lines is a relatively easy task in Galactic H~{\sc ii} regions and PNe. However, their emissivity decreases rapidly with metallicity and with decreasing surface brightness of the objects so, detecting them is a challenging task in the extragalactic domain, especially in objects beyond the Local Group. However, the combination of high-sensitivity spectrographs with large aperture (10m type) telescopes have allowed the detection of the auroral [O~{\sc iii}] $\lambda$4363 \AA\ line in at least 18 star-forming galaxies at  $z>1$ (see \cite{sandersetal19} and references therein).

On the other hand, line ratios sensitive to $n_e$ come from levels with very similar energy, so that the ratio of their populations does not depend on $T_e$. These levels show different transition probabilities or different collisionally de-excitation rates, such that the ratio between the emission lines generated is strongly dependent on the electron density of the photoionized gas. Typical optical density diagnostics are: [O~{\sc ii}] $\lambda$3726/$\lambda$3729, [S~{\sc ii}] $\lambda$6716/$\lambda$6731, [Cl~{\sc iii}] $\lambda$5517/$\lambda$5537 and [Ar~{\sc iv} $\lambda$4711/$\lambda$4741.

A precise determination of the physical conditions is crucial to derive reliable abundances from CELs. As the abundances are computed relative to H by using the relative intensities of CELs or ORLs relative to a H~{\sc i} ORL (usually H$\beta$), and given the very different dependence of the emissivity of CELs and ORLs (see \cite{peimbertetal17}) the abundances from CELs show a strong (exponential) dependence on $T_e$, while abundances computed from faint metallic ORLs are almost $T_e$ dependent. This has important implications for the so-called abundance discrepancy problem (see Section~\ref{sec:adf}). 

Once physical conditions are computed one has to decide the temperature and density structure that is going to be assumed in the nebula. The most common approach, is to assume a two-zone scheme, where the high ionization zone is characterized by $T_e$(high) (usually $T_e$([O~{\sc iii}]), the low-ionization zone is characterized by $T_e$(low) (usually $T_e$([N~{\sc ii}])  and the density is considered homogeneous in the whole nebula and is characterized by $n_e$([S~{\sc ii}]). Then, each temperature is applied to compute ionic abundances of species with similar ionization potentials than the ion used in the $T_e$ diagnostic. In a typical spectra, $T_e$(low) is applied to compute abundances of N$^+$, O$^+$, S$^+$, Cl$^+$ and Fe$^+$, while $T_e$(high) is used for the remaining ionic species observed in the optical spectra.

However, recent results from \cite{dominguezguzmanetal19} have shown that this scheme can be erroneous. These authors, from deep, high-resolution spectra of H~{\sc ii} regions in the Magellanic Clouds, have proposed that for some ions, it is better to adopt other scheme in order to avoid trends with metallicity. In particular, they propose to use $T_e$([N~{\sc ii}]) to calculate Cl$^{2+}$ and the mean of $T_e$([N~{\sc ii}]) and $T_e$([O~{\sc iii}]) for S$^{2+}$ and Ar$^{2+}$, finding that, in such cases, Cl/O, S/O and Ar/O are approximately constant with metallicity (see their Fig.~3) as expected for $\alpha$-elements.

In deep spectra, covering the whole optical (or even up to 1$\mu$m) wavelength range more electron temperature and density diagnostics will be available. In such cases, $T_e$(low) and $T_e$(high) can be computed as the average of the values obtained from different diagnostics, which are generally in reasonable good agreement within the uncertainties (see e.~g. \cite{tsamisetal03, garciarojasetal06, wessonetal05}). In some cases, particularly in relatively high-density PNe ($n_e>10^4$cm$^{-3}$), density stratification can be observed, with the [Ar~{\sc iv}] densities being larger than those computed with the other three diagnostics (see \cite{wangetal04}). In such cases it is better to consider also a two-zone density model (see \cite{garciarojasetal13}). 
In some extreme cases of extremely young and dense PNe, with densities higher than the critical densities of the upper levels of the transitions producing the [Ar~{\sc iv}] lines, all the classical electron density diagnostics will be saturated and can provide inaccurate densities. An alternative density indicator is based on the analysis of [Fe~{\sc iii}] emission lines, which are robust density diagnostics when collisional de-excitation dominates over collisional excitation. Indeed, if inappropriate density diagnostics are used, then physical conditions deduced from 
commonly used line ratios will be in error, leading to unreliable chemical abundances for these
objects. (see \cite{mesadelgadoetal12}).

\subsubsection{Analysis tools}
\label{sec:analysis}

The first public code for the computation of physical conditions and ionic abundances was {\sc fivel} \cite{derobertisetal87}, an interactive {\sc FORTRAN} program which used a basic five-level atom approximation, which considers that only the five low-lying levels (i.~e. at energies $\leq$5 eV  above the ground state are physically relevant for computing the observed emission line spectrum. Later, \cite{shawetal95} developed {\sc nebular}, a set of software tools (based in the {\sc fivel} program, but extending it to an N-level atom) in the {\sc iraf/stsdas}\footnote{{\sc iraf} is distributed by National Optical Astronomy Observatories, which is operated by AURA (Association of Universities for Research in Astronomy), under cooperative agreement with NSF (National Science Foundation).} environment that allow the user to compute diagnostic for a variety of ground-state electron configurations, and compute ionic abundances separately for up to 3 zones of ionization. The main advantage of {\sc nebular} is that it can be scripted. However, changes of atomic data sets is not trivial and computations of elemental abundances are not included.

(\cite{wessonetal12}) developed the  Nebular Empirical Analysis Tool ({\sc neat}\footnote{https://www.nebulousresearch.org/codes/neat/}), a very simple to use code written in {\sc FORTRAN90} which requires little or no user input to return robust results, trying to provide abundance determinations as objective as possible. One of the main advantages of this code is that it can evaluate uncertainties of the computed physical conditions and abundances by using a Monte Carlo approach. Another advantage of this code is that it also accounts for the effect of upward biasing on measurements of lines with low signal-to-noise ratios, allowing to reduce uncertainties of abundance determinations based on these lines. Finally, as atomic data for heavier elements than helium are stored externally in plain text files, the user can easily change the atomic data. 

The last package to be offered in the field has been {\sc PyNeb}\footnote{https://github.com/Morisset/PyNeb\textunderscore devel} (\cite{luridianaetal15}) which is completely written in python and is designed to be easily scripted, and is more flexible and therefore, powerful than its predecessors. This package allow the user to easily change and update atomic data as well as providing tools to plot and compare atomic data from different publications.

\subsection{Abundances in distant photoionized nebulae: the strong line methods}
\label{sec:strong}

In the absence of reliable plasma diagnostics (a common fact in extragalactic objects) in giant H~{\sc ii} regions or integrated spectra of galaxies, one needs to use alternative methods to derive accurate chemical abundances. This is especially important to estimate the metallicities of giant extragalactic H~{\sc ii} regions as well as of local and high-redshift emission-line galaxies and hence, it has a relevant influence on the study of the chemical evolution of the Universe. 

The first mention of the strong-line methods was 40 years ago, when \cite{pageletal79} and \cite{alloinetal79} proposed a method to compute the oxygen abundance using strong lines only: the $R_{23}$ method, in which oxygen abundance is a one dimensional  function of the $R_{23}$ parameter, defined as:

\begin{equation}
R_{23}=\frac{[{\rm O II}]\lambda 3727+[{\rm O III}]\lambda 4959+5007}{H\beta}
\end{equation}

This method was calibrated using the few relevant photoionization models available at that time.
The problem with dealing with $R_{23}$ is that it is double valued with respect to metallicity. In fact, at low oxygen  abundances --12+log(O/H) $\lesssim$ 8.0-- the $R_{23}$ index increases with the abundance, while for high oxygen  abundances --12 + log(O/H) $\ge$ 8.25-- the efficiency of the cooling caused by metals make $R_{23}$ to drop with rising abundance. There is also a transition zone between 8.0 and 8.25 (see e.~g. \cite{pilyuginthuan05} for a  detailed description of the high and low metallicity branches). This method has been refined multiple times since then and several calibrations, using data sets with abundances from the direct method (e.~g. \cite{pilyuginthuan05, pilyuginetal10}), using photoionization model grids (e.~g. \cite{mcgaugh91, kewleydopita02}), or a combination of both, are now available in the literature. An overview of the most popular calibrations of strong-line methods can be found in \cite{lopezsanchezesteban10}. A comprehensive critical evaluation of the different semi-empirical strong-line methods has been done by \cite{lopezsanchezetal12} who also develop a method for reducing systematics in the techniques to compute chemical abundances by using electron temperatures and ionization correction factors.

In the last years, mainly thanks to the increasingly easy access to super-computing resources, new approaches have been proposed. Bayesian methods have been used by several authors to determine chemical abundances in extragalactic targets (e.~g. \cite{valeasarietal16}) although the priors should be selected cautiously to avoid unreliable results. On the other hand, as in most of astronomy fields, machine learning techniques are also being used to infer chemical abundances (see e.~g. \cite{ho19}). However, as has been pointed out by \cite{stasinska19}, making use of an illustrative example, the use of these techniques ignoring the underlying physics can lead to unphysical inferences.

\cite{stasinska19} have argued in a comprehensive review that although strong-line methods are routinely used to estimate metallicities owing to their apparent simplicity, the users need to have a solid background on the physics of H~{\sc ii} regions to understand the approximations made on the different approaches, and the limitations each calibration has, to avoid biases, misinterpretations and mistakes. 

Even taking into account the drawbacks mentioned above, strong-line methods have been widely used for studying giant H~{\sc ii} regions and emission line galaxies in large long-slit spectroscopic surveys as the Sloan Digital Sky Survey (SDSS) \cite{stasinska06}, or 2D spectroscopic surveys as MANGA (e.~g. \cite{parikhetal19, sanchezmenguianoetal19}), CALIFA (e.~g. \cite{sanchezetal14}), and AMUSING (e.~g. \cite{sanchezmenguianoetal18}). 

\section{Advances in abundances determinations in photoionized nebulae}
\label{sec:abund}

In this section I will focus on the latest advances that have been reached in the field of photoionized nebulae. I will pay special attention to atomic data, ionization correction factors and the abundance discrepancy problem, that have been traditionally claimed as potential sources of uncertainty in chemical abundance determinations.

\subsection{Atomic data}
\label{sec:atomic}

The atomic data used for computing abundances in photoionized nebulae are ususally considered as a black box by the users. Most users consider the default atomic data sets used by their favourite analysis tools or directly use the last available atomic data in the literature for each ion. In the last years large compilations of atomic data have been done in the {\sc chianti}\footnote{http://www.chiantidatabase.org} and {\sc nist}\footnote{http://physics.nist.gov} databases, although the available atomic data in each database do not always match for a given ion. \cite{luridianaetal11, luridianagarciarojas12} discussed how to ensure that atomic data are correctly understood and used, as well as on the typical uncertainties in atomic data.

High-quality observations of photoionized nebulae are a powerful tool to check the reliability of atomic data. \cite{copettiwritzl02} and \cite{wangetal04} found, using a large data set of PNe spectra and comparing electron density estimates for PNe based on different density diagnostics, that the [O~{\sc ii}] transition probabilities calculated by \cite{wieseetal96} yielded systematically lower electron densities than those computed using the [S~{\sc ii}] diagnostic, and that such discrepancies were caused by errors in the computed transition probabilities. Moreover, \cite{wangetal04} found that the transition probabilities of \cite{zeippen87} and the collision strengths of \cite{mclaughlinbell98} were completely inconsistent with observations at the high and low density limits, respectively, and should be ruled out.

\cite{juandediosrodriguez17} determined chemical abundances of O, N, S, Ne, Cl and Ar for a sample of PNe and H~{\sc ii} regions and evaluated the impact of using different sets of atomic data on the computed physical conditions and abundances. These authors used all the possible combinations of 52 different sets of transition probabilities and collision strengths to calculate physical conditions and chemical abundances, finding that different combinations of atomic data introduce differences in the derived abundances that can reach or surpass 0.6-0.8 dex at higher densities ($n_e>10^{-4}$ cm$^{-3}$ in several abundance ratios like O/H and N/O. Removing the data sets that introduce the largest differences can reduce the total uncertainties, although they can still remain in high-density objects. Additionally, they have pointed out that special attention should be paid to the transition probabilities of the S$^+$, O$^+$, Cl$^{++}$ and Ar$^{3+}$ density diagnostic lines, and to the collision strengths of Ar$^{3+}$ which, if incorrectly selected, can lead to unreliable chemical abundances in high-density nebulae.

Finally, \cite{stasinska19} has pointed out that the role of atomic data in strong-line method calibrations cannot be ignored. Recent changes in routinely used atomic data have revealed that they play a crucial role in direct abundance determinations and in photoionization models.

\subsection{Ionization correction factors}
\label{sec:icf}

The elemental abundance of a particular element is computed by adding up the ionic abundances of all the ions present in a nebula. However, it is usually found that not all the ions of a given element are observed, whether because they are emitted in a different spectral range than that observed or because the spectra is not deep enough to detect them. Therefore, the contribution of these unobserved ions should be estimated in someway. With this aim, the use of Ionization Correction Factors (ICFs) was proposed by \cite{peimbertcostero69}. These authors proposed to use similarities between ionization potentials of different ions to construct ICFs. This approach has been used by several authors since then (see e.~g. \cite{peimberttorrespeimbert77}). However, \cite{stasinska02} argued that these ICFs should be treated with caution because the ionization structure in a photoionized nebula does not depend only on the ionization potential. Moreover, it has been shown that using
recent photoionization models, these simple expressions are not
always valid and new ICFs are needed to obtain more reliable abundances (see e.~g. \cite{delgadoingladaetal19})

The alternative is to compute ICFs using photoionization models, where the physics involved in ionized nebulae is treated with much more detail. Photoionization models allow to compute the detailed ionization structure of the various elements present in a
nebula, by taking into account all the processes that govern
ionization and recombination (i.e. mostly photoionization, radiative
and di-electronic recombination, and charge exchange), as well as
all the heating and cooling processes that determine the electron
temperature (\cite{delgadoingladaetal14}).  

Traditionally, different ICFs have been computed for H~{\sc ii} regions and PNe, given the differences in the hardness of the radiation field and the different ionic species detected in each type of object (see Section~\ref{sec:intro}). Several authors have derived ICFs from photoionization models for H~{\sc ii} regions (\cite{stasinska78, mathisrosa91, izotovetal94, garnettetal99, izotovetal06, perezmonteroetal07, dorsetal16}) and for PNe (\cite{kingsburghbarlow94, rodriguezrubin05, delgadoingladaetal14}). It is not the scope of this text to show the details of the different approaches used to compute ICFs from photoionization models, but I think it is worth mentioning some of the most widely used ICF schemes. \cite{izotovetal06} re-evaluated empirical expressions for the abundance determination of N, O, Ne, S, Cl, Ar and Fe to compute abundances of emission-line galaxies from the Data Release 3 of the Sloan Digital Sky Survey (SDSS). They took special care in the selection of atomic data and constructed an appropriate grid of photoionization models with state-of-the art model atmospheres. In particular, these authors take care of a problem that should not be ignored in the computation of photoionization models, which is the uncertain rate of the dielectronic recombination for sulfur, chlorine and argon ions. They compared the abundances of these elements calculated with different assumed dielectronic recombination rates and could put some constraints on these rates. Additionally, following an approach that was defined by \cite{stasinskaizotov03} these authors proposed different ICFs depending on the metallicity range of the nebulae. 
Regarding PNe, \cite{delgadoingladaetal14} constructed ICFs for He, N, O, C, Ne, S, Cl, and Ar using a large grid of photoionization models that are representative of most of the observed PNe. Besides the obvious advantage of covering a wide range of physical parameters with a large photoionization model grid, the main advantage of this work is the provision of analytical expressions to estimate the uncertainties arising from their computed ICFs.  

Finally, a third scheme to compute ICFs is to derive analytical expressions obtained from observational fittings to large sets of high-quality data (see e.~g. \cite{estebanetal15} for Cl, and \cite{bergetal19} for C).

In section~\ref{sec:ncap} I will come back to the ICFs mentioning some works devoted to the computation of ICFs for neutron-capture elements in PNe. 

\subsection{Oxygen enrichment in PNe}
\label{sec:oxygen}

Oxygen is the element for which more reliable abundances can be obtained and, therefore, it has been traditionally used as a proxy for metallicity.  In H~{\sc ii} regions, oxygen reflects the current abundance in the ISM, while in PNe, it is supposed to reflect the chemical composition of the environment where the star was born because its abundance remain unchanged during the life of the star (\cite{delgadoingladaetal15}). However, AGB stars can modify the oxygen abundance by two mechanisms: the third dredge-up (TDU) and the hot bottom burning (HBB), although only nucleosynthesis models which include extra-mixing processes like diffusive convective overshooting (e.~g. \cite{marigo01, garciahernandezetal16, pignatarietal16}) predict a significant production of oxygen.

\begin{figure}  
\sidecaption
\hspace{0.25cm}
\includegraphics[width=\textwidth]{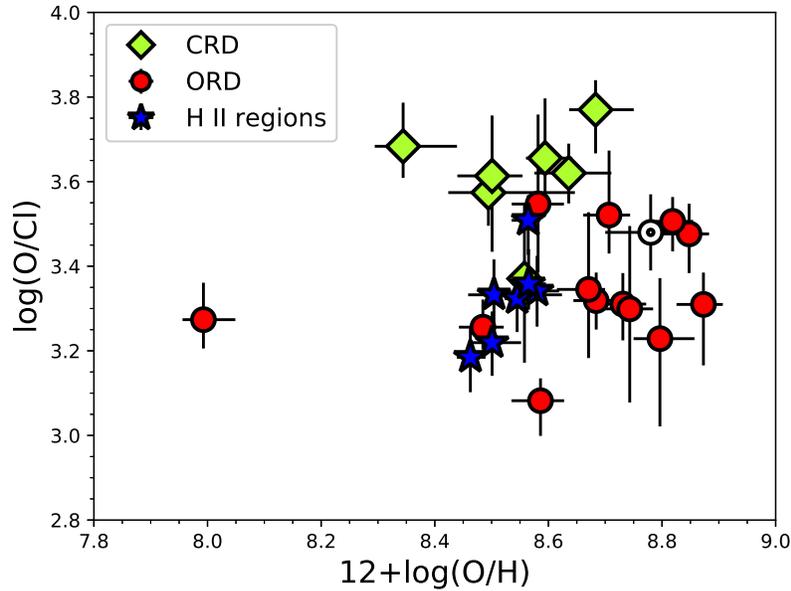}
\caption{Values of O/Cl as a function of O/H for a sample of Galactic PNe and H~{\sc ii} regions (see \cite{delgadoingladaetal15}). The red circles represent PNe with oxygen-rich dust, the green diamonds PNe with carbon-rich dust, and the blue stars the H~{\sc ii} regions. The protosolar abundances of \cite{lodders10} are overplotted with the solar symbol. Plot made with data gently provided by Gloria Delgado-Inglada.}
\label{fig:5}
\end{figure}

Until recently, the only observational probes of oxygen production in AGB stars have been restricted to low-metallicity PNe  (see e.~g. \cite{leisydennefeld06}). However, using deep, high-quality optical spectra (with spectral resolution better than 1 \AA) \cite{delgadoingladaetal15} recomputed accurate abundances of He, O, N, Ne, C, Ar, and Cl in 20 PNe and 7 H~{\sc ii} regions in our Galaxy at near-solar metallicities. These authors found that all but one of the Galactic PNe with C-rich dust (the one with the highest metallicity according to Cl/H) show higher O/Cl values than the PNe with O-rich dust and the H~{\sc ii} regions (see Fig.~\ref{fig:5}), and interpret that result as O is enriched in C-rich PNe due to an efficient third dredge-up in their progenitor stars. These results have been confirmed later by nucleosynthesis models including convective overshooting by \cite{garciahernandezetal16, garciahernandezetal16b}.

These findings confirm that oxygen is not always a good proxy of the original ISM metallicity and other chemical elements such as chlorine or argon, the abundance of which is unaltered in the evolution of low- and intermediate-mass stars, should be used instead. Additionally, as has been pointed out by \cite{garciahernandezetal16b}, the production of oxygen by low-mass stars should be thus considered in galactic-evolution models.

\subsection{Abundance gradients in the Milky Way and in nearby spiral galaxies from direct abundance determinations}
\label{sec:grad}

The study of the radial distribution of the gas phase metallicity in a Galaxy (usually using oxygen as a proxy for the metallicity) is fundamental for our understanding of the evolution of Galaxies. The pioneering studies on the gradient of abundances in spiral galaxies were those of \cite{searle71} and \cite{pageledmunds81}, which were based on the spectral differences found by \cite{aller42} between the H~{\sc ii} regions in the spiral galaxy M33. \cite{shaveretal83} were the first in carrying out an homogeneous study of abundance gradients in the Milky Way (hereinafter, MW) with a relatively large sample of H~{\sc ii} regions. However, these authors rely on electron temperatures determined from radio RLs and abundances from optical lines, obtaining a relatively large scatter at a given Galactocentric distance. Since these pioneering works, several authors have computed radial abundance gradients using the direct method in our Galaxy (e.~g. \cite{deharvengetal00, estebanetal05, rudolphetal06, estebanetal17, fernandezmartinetal17}) and in external galaxies (see e.~g. \cite{garnettetal97, zuritabresolin12, croxalletal15, bergetal15, toribiosanciprianoetal16} and references therein). 

Regarding the MW, the relatively large scatter at a given Galactocentric distance found in several works has been claimed as a possible indication that the gas is not as well mixed as commonly thought (see e.~g. \cite{rudolphetal06}). Moreover, \cite{balseretal11} found significant differences in the radial gradient of O in the MW depending on the Galactic azimuth region considered, strengthening the idea that metals are not well mixed at a given radius. However, \cite{estebangarciarojas18} made an homogeneous analysis using a set of deep optical spectra of 35 H~{\sc ii} regions, from which they computed accurate physical conditions and ionic and elemental abundances, finding that the scatter of the N and O abundances of H~{\sc ii} regions is of the order of the observational uncertainties, indicating that both chemical elements seem to be well mixed in the ISM at a given Galactocentric distance (see the comparison between radial O abundance from this work and that of \cite{rudolphetal06} in Fig.~\ref{fig:5}).

In the extragalactic domain, it is worth mentioning the existence of the CHemical Abundances of Spirals (CHAOS) project, which is devoted to surveying several spiral galaxies to determine precise ``direct'' abundances in large samples of H~{\sc ii} regions in spiral galaxies (see \cite{bergetal15, croxalletal15, croxalletal16}). This project has increased by more than an order of magnitude the number of H~{\sc ii} regions with direct measurements of the chemical abundances in nearby disk galaxies (see \cite{bergetal15}).

There are many open problems with the abundance gradients of the MW and nearby spiral Galaxies such as a possible temporal evolution (\cite{macielcosta13, stanghellinihaywood18}) based on the differences found in the gradients using different populations of PNe and H~{\sc ii} regions; the existence or not of a flattening of the gradient in the outer disc of spiral galaxies, including the MW (\cite{henryetal10, estebanetal17, sanchezmenguianoetal18}) or in the inner disc (\cite{estebanetal17, sanchezmenguianoetal18}); distance determinations uncertainties, particularly for PNe (see \cite{stanghellinihaywood18}) or, as mentioned in section~\ref{sec:oxygen}, the applicability of oxygen as a reliable element to trace the metallicity in PNe (\cite{delgadoingladaetal15}). Some of the limitations that, in my opinion should be taken into account have been summarized in \cite{garciarojas18}.

The determination of precise radial metallicity gradients are precious constraints for chemical evolution models of the MW in particular and of spiral galaxies in general. The presence of a negative gradient agrees with the the stellar mass growth of galaxies being inside-out (see e.~g. \cite{perezetal13}). However, additional information, such as the possible temporal evolution of the gradients (see e.~g. \cite{macielcosta13, stanghellinihaywood18}) can give information about physical processes that can modify gradients, and that should be considered by chemical evolution models (see discussion by \cite{stanghellinihaywood18} and references therein). As H~{\sc ii} regions reflect the young stellar populations and, on the other hand PNe, reflect older stellar populations (with a relatively large spread in ages) a careful comparison between gradients obtained from different objects is very useful to constrain the temporal evolution of the gradient predicted by chemical evolution models (\cite{mollaetal19}).

\subsection{The abundance discrepancy problem}
\label{sec:adf}

The abundance discrepancy problem is one of the major unresolved problems in nebular astrophysics and it has been around for more than seventy years (\cite{wyse42}). It consists in the fact that in photoionized nebulae $-$both H~{\sc ii} regions and PNe$-$ ORLs provide abundance values that are systematically larger than those obtained using CELs. Solving this problem has obvious implications for the measurement of the chemical content of nearby and distant galaxies, because this task is most often done using CELs from their ionized ISM.

For a given ion, the abundance discrepancy factor (ADF) is defined as the ratio between the abundances obtained from ORLs and CELs, i. e.,

\begin{equation}
ADF(X^{i+}) = (X^{i+}/H^+)_{ORLs}/(X^{i+}/H^+)_{CELs}, 
\end{equation}

and is usually between 1.5 and 3, with a mean value of about 2.0 in H~{\sc ii} regions and the bulk of PNe (see e.g. \cite{garciarojasesteban07, mcnabbetal13}, but in PNe it has a significant tail extending to much larger values, up to 2--3 orders of magnitude\footnote{An updated record of the distribution of values of the ADF in both H~{\sc ii} regions and PNe can be found in Roger Wesson's webpage: http://nebulousresearch.org/adfs}. It is important to remark that the ADF is most easily determined for O$^{2+}$ owing to both CELs and RLs are straightforward to detect in the optical. ADFs can be also determined for other ions, such as C$^{2+}$, N$^{2+}$, and Ne$^{2+}$, although the obtained values are more uncertain because CELs and RLs are detected in different wavelength ranges (in the case of C$^{2+}$ and N$^{2+}$) or because RLs are intrinsically very faint (in the case of Ne$^{2+}$). 

The possible origin of this discrepancy has been discussed for many years and three main scenarios have been proposed: 

\begin{itemize}
\item \cite{peimbert67} was the first proposing the presence of temperature fluctuations in the gas to explain the discrepancy between $T_e$([O~{\sc iii}]) and $T_e$(H~{\sc I}) derived from the Balmer jump. After that seminal work, \cite{peimbertcostero69} developed a scheme to correct the abundances computed from CELs for the presence of temperature inhomogeneities. Later, \cite{torrespeimbertetal80} suggested that the discrepancy between ORLs and CELs abundances could be explained if spatial temperature variations over the observed volume were considered. \cite{peimbertetal17} have recently summarized the mechanisms proposed to explain and maintain the presence of temperature fluctuations in photoionized nebulae.

\item \cite{torrespeimbertetal90} were the first in proposing the existence of chemical inhomogeneities in the gas as a plausible mechanism to explain the abundance discrepancy. This scenario was later expanded by \cite{liuetal00}, who claimed that metal-rich (i.~e. H-poor) inclusions could be the clue to resolve the abundance discrepancy problem; in this scenario, metal ORLs would be emitted in the metal-rich inclusions, where cooling has been enhanced, while CELs would be emitted in the ``normal'' metallicity (H-rich) zones. This model was tested by several authors by constructing two-phase photoionization models that, in several cases, successfully simultaneously reproduced the ORLs and CELs emissions in H~{\sc ii} regions \cite{tsamispequignot05} and PNe \cite{yuanetal11}. However, at the present time, the origin of such metal-rich inclusions remains elusive, although some scenarios have been proposed for PNe (\cite{henneystasinska10}) and H~{\sc ii} regions (\cite{stasinskaetal07}). In the last years, increasing evidence has been found of a link between the presence of a close central binary star at the heart of PNe and very high ($>$10) ADFs (see Section~\ref{sec:binary}).

\item A third scenario was brought into play by \cite{nichollsetal12}, who proposed that the departure of the free electron energy distribution from the Maxwellian distribution ($\kappa$-distribution) could explain the abundance discrepancies owing to the presence of a long tail of supra-thermal electrons that contribute to an increase in the intensity of the CELs at a given value of the kinetic temperature.  However, in the last years little theoretical (\cite{mendozabautista14, ferlandetal16, drainekreisch18}) or observational (\cite{zhangetal16}) support has been presented for this scenario in photoionized nebulae. \cite{ferlandetal16} have shown that the heating or cooling timescales are much longer than the timescale needed to thermalize supra-thermal electrons because they can only travel over distances that are much shorter than the distances over which heating rates change, implying that the electron velocity distribution will be close to a Maxwellian one  long before the supra-thermal electrons can affect the emission of CELs and RLs. Moreover, \cite{drainekreisch18} demonstrated analytically that the electron energy distribution relaxes rapidly to a steady-state distribution that is very close to a Maxwellian, having negligible effects on line ratios. 
\end{itemize}

One of the most active groups in the study of the abundance discrepancy in PNe in the last two decades has been the University College London/U. Beijing group, who have developed deep medium-resolution spectrophotometry of dozens of PNe to compute the physical and chemical properties of these objects from ORLs. 
In one of the most detailed and comprehensive studies of this group, \cite{wangliu07} showed that the values of the ADF deduced for the four most abundant second-row heavy elements (C, N, O and Ne) are comparable (see their Fig.~18). However, they also computed abundances from ORLs from a third-row element (Mg) and they found that no enhancement of ORL abundances relative to CEL ones is obvious for Mg: the average Mg abundances from  ORLs for disk PNe remained in a range compatible to the solar photospheric value, even taking into account the small depletion expected for this element onto dust grains (less than 30\%). 
Finally, these authors also showed that, regardless of the value of the ADF, both CEL and ORL abundances yield similar relative abundance ratios of heavy elements such as C/O, N/O and Ne/O . This has important implications, especially in the case of the C/O ratio, given the difficulties of obtaining this ratio from UV CELs (see Sect.~\ref{sec:co_ratio}).

Several authors have strongly argued in favour of the inhomogeneous composition of PNe and against pure temperature fluctuations (see e.~g. \cite{liuetal06} and references therein); some of the reasoning that has been presented supporting this model are: i) far-IR [O~{\sc iii}] CELs, which in principle, have a much lower dependence on electron temperature than optical CELs, provide abundances that are consistent with those derived from optical CELs (see e.~g. \cite{liuetal00}); ii) the analysis of the physical conditions using H, He, O and N ORLs yields electron temperatures that are much lower than those computed from classical CEL diagnostic ratios (see \cite{tsamisetal04, zhangetal04, zhangetal09, fangliu11}); additionally, ORL density diagnostics provide densities that are higher than those derived from CEL diagnostics; iii) chemically homogeneous photoionization models do not reproduce the required temperature fluctuations to match CEL and ORL abundances, while bi-abundance photoionization models including an H-poor (i.~e. metal-rich) component of the gas successfully reproduce the observed intensities of both CELs and ORLs (e.~g. \cite{yuanetal11}). All these arguments strongly favour the presence of a low-mass component of the gas that is much colder and denser than the ``normal'' gas, and that is responsible for the bulk of the ORL emission. 
However, we cannot rule out the possibility that different physical phenomena can contribute simultaneously to the abundance discrepancy in PNe.

Some physical phenomena have been proposed to explain the abundance discrepancy in the framework of temperature fluctuations or chemical inhomogeneities scenarios (\cite{henneystasinska10, peimbertetal17}). However, until very recently, there was no observational proof that demonstrated a single physical process to be responsible for the abundance discrepancy. 
Some recent works on the Orion nebula have observationally linked the abundance discrepancy to the presence of high velocity flows (\cite{mesadelgadoetal09}) or to the presence of high density clumps, such as proto-planetary disks (\cite{tsamisetal11, mesadelgadoetal12}. On the other hand, \cite{liuetal06} found a very extreme value of the ADF for the PN Hf\,2--2 (ADF$\sim$70) and, for the first time, suggested the possibility that this large ADF could be related to the fact that the central star of the PN, which is a close-binary star, has gone through a common-envelope phase. 

\subsubsection{The link between close binary central stars and large abundance discrepancy factors}
\label{sec:binary}

Several papers in recent years have confirmed the hypothesis proposed by \cite{liuetal06} that the largest abundance discrepancies are reached in PNe with close-binary central stars. \cite{corradietal15} found that three PNe with known close-binary central stars showed high ADFs, with the PN Abell~46, with an ADF(O$^{2+}$)$\sim$120, and as high as 300 in its inner regions, being the most extreme object. Their spectroscopic analysis supports the previous interpretation that, in addition to ``standard'' hot (T$_e$$\sim$10$^4$~K) gas, a colder (T$_e$$\sim$10$^3$~K), metal-rich, ionized component also exists in these nebulae.  Both the origin of the metal-rich component and how the two gas phases are mixed in the nebulae are basically unknown. Moreover, this dual nature is not predicted by mass-loss theories. However, it seems clear that the large-ADF phenomena in PNe is linked to the presence of a close-binary central star. In fact, \cite{wessonetal18} have recently completed a survey of the ADFs in seven PNe with known close-binary central stars and they found ADFs larger than 10 for all of them, confirming the strong link between large ADFs and close-binary central stars.
On the other hand, several spectroscopic studies have shown that the ORL emitting plasma is generally concentrated in the central parts of the PNe. This occurs in PNe with known close-binary central stars and large ADFs (e.~g. \cite{corradietal15, jonesetal16}), in PNe with low-to-moderate ADFs and no indication of binarity (e.~g. \cite{liuetal01,garnettdinerstein01}) and in PNe with relatively large ADFs but no known close-binary central star (e.~g. M\,1--42, see \cite{garciarojasetal17}).

\begin{figure}  
\sidecaption
\hspace{0.25cm}
\includegraphics[width=\textwidth]{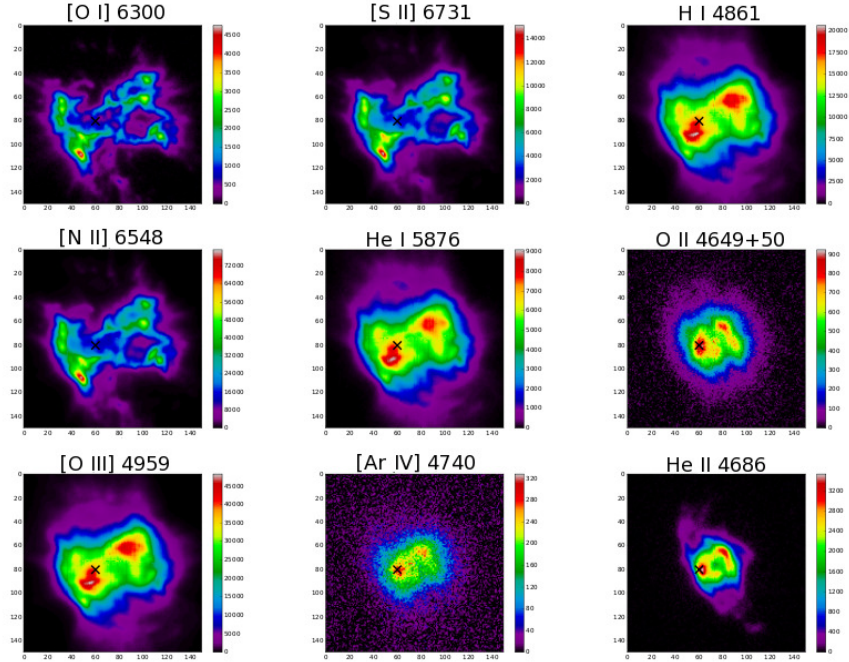}
\caption{MUSE Emission line maps of several lines in the PN NGC\,6778, ordered by ionization potential of the parent ion: from left to right and from top to bottm: [O~{\sc i}] $\lambda$6300 \AA, [S~{\sc ii}] $\lambda$6731 \AA, H$\beta$ $\lambda$4861 \AA, [N~{\sc ii}] $\lambda$6548 \AA, He~{\sc i} $\lambda$5876 \AA, O~{\sc ii} $\lambda\lambda$4649+50 \AA ORLs, [O~{\sc iii}] $\lambda$4959 \AA CEL, [Ar~{\sc iv}] $\lambda$4740 \AA and He~{\sc ii} $\lambda$4686 \AA. It is worth re-emphasising that the O~{\sc ii} and the [O~{\sc iii}] emission comes from the same ion: O$^{2+}$ . The ``x'' marks a reference spaxel}
\label{fig:6}
\end{figure}

\cite{garciarojasetal16} recently obtained the first direct image of the PN NGC\,6778 (a PN with ADF$\sim$20) in O~{\sc ii} recombination lines, taking advantage of the tunable filters available at the OSIRIS instrument in the 10.4m Gran Telescopio Canarias (GTC). They found that in NGC~6778, the spatial distribution of the O~{\sc ii} $\lambda\lambda$4649+50 ORL emission does not match that of the [O~{\sc iii}] $\lambda$5007 CEL. \cite{garciarojasetal17} found the same behaviour in Abell~46 using direct tunable filter images centred at $\lambda\lambda$4649+51 \AA.

Moreover, \cite{garciarojasetal17} presented preliminary results obtained from deep 2D spectroscopic observations with MUSE at the 8.2m Very Large Telescope (VLT) of five southern large-ADF PNe, and they confirmed this behaviour in at least the PNe Hf~2-2 (ADF$\sim$84), M~1-42 and NGC\,6778 (both with ADF$\sim$20). In Fig~\ref{fig:6} we show the MUSE emission line maps of several emission lines in the PN NGC\,6778. The emission maps are ordered by increasing ionization potential of the parent ion from left to right and from top to bottom. It is clear that O~{\sc ii} $\lambda$4649+50 ORLs emission is more centrally concentrated that [O~{\sc iii}] $\lambda$4959 CEL emission, and seems to be emitted in a zone that correspond to a higher ionization specie. A similar result has been found by \cite{richeretal13} from a kinematical analysis of several heavy metal ORLs and CELs in NGC\,7009. These authors found that the kinematics of ORLs and CELs were discrepant and incompatible with the ionization structure of the nebula, unless there is an additional plasma component to the CEL emission that arises from a different volume from that giving rise to the RL emission from the parent ions within NGC\,7009. Similarly, \cite{richeretal17} found that the kinematics of the C~{\sc ii} $\lambda$6578 line is not what expected if this line arises from the recombination of C$^{2+}$ ions or the fluorescence of C$^+$ ions in ionization equilibrium in a chemically homogeneous nebular plasma, but instead its kinematics are those appropriate for a volume more internal than expected.

These results clearly support the hypothesis of the existence of two separate plasmas, at least in these large-ADF PNe, with the additional indication that they are not well mixed, perhaps because they were produced in distinct ejection events related to the binary nature of the PN central star. \cite{wessonetal18} propose that a nova-like outburst from the close-binary central star could be responsible for ejecting H-deficient material into the nebulae soon after the formation of the main nebula. This material would be depleted in H, and enhanced in C,N, O, and Ne, but not in third row elements. It is worth mentioning the similarity of these plasma component with some well-known old nova shells as CP Pup and DQ Her that show low $T_e$ and strong ORLs (\cite{williamsetal78,  williams82}).

\subsection{The C/O ratio from recombination lines}
\label{sec:co_ratio}

The determination of accurate C/H and C/O ratios in H~{\sc ii} regions is of paramount importance to constrain chemical evolution models of galaxies (\cite{carigietal05}). Furthermore, C is an important source of opacity in stars and one of the main elements found in interstellar dust and organic molecules, making it one of the biogenic elements (\cite{estebanetal14}). Despite its importance, C abundances in H~{\sc ii} regions have been poorly explored as they are traditionally derived from UV observations of the semi-forbidden C~{\sc iii}] $\lambda$1909 and C~{\sc ii}] $\lambda$2326 CELs from space (\cite{bergetal16, bergetal19}). For PNe the situation is better, owing to several successful $IUE$ and $HST$ programs that have provided reliable measurements of these lines in dozens of objects (see e.~g. \cite{liuetal04, stanghellinietal09, dufouretal15, henryetal18, milleretal19} and references therein). 
However, the determination of reliable fluxes from these lines is difficult as they are severely affected by interstellar reddening. Moreover, the emissivities of these lines are also very dependent on the electron temperature. Finally, aperture effects owing to the different areas covered by UV and optical observations must be taken into account to guarantee the observation of the same volume of the nebula in both ranges. Alternatively, scanning techniques have been used to match UV International Ultraviolet Explorer (IUE) observations and optical spectroscopy (see e.~g. \cite{liuetal00, liuetal01}).

The determination of C and O abundances from ORLs skips these difficulties and takes advantage of the fact that using the same type of lines (CELs or ORLs) the C/O ratios remain the same in most of the objects were both types of ratio have been computed (\cite{wangliu07, delgadoingladarodriguez14}. Thanks to the new CCDs with improved efficiency in the blue and the use of large telescopes, several high-quality observations of the C~{\sc ii} $\lambda$4267 ORL in PNe have been achieved in the last years (e.~g.  \cite{peimbertetal04, liuetal04, wessonetal05, sharpeeetal07, wangliu07, garciarojasetal12, fangliu11, garciarojasetal18}. 

\subsubsection{C/O ratios in H~{\sc ii} regions}

\cite{estebanetal05} computed for the first time the C/H and C/O radial gradients of the ionized gas from ORLs in the MW. Later, \cite{estebanetal09} derived these gradients in M101, and \cite{toribiosanciprianoetal16} in M33 and NGC\,300. 

The general conclusion of these works is that C abundance gradients are always steeper than those of O, producing negative C/O gradients across the galactic disks which reflect the non-primary behavior of C enrichment. Furthermore, the comparison between the C/H and C/O gradients obtained in the MW with state-of-the-art chemical evolution models revealed that the obtained C gradients can only been explained if the C produced by massive and low-intermediate mass stars depends strongly on time and on the Galactocentric distance (\cite{carigietal05}).

From the C/O ratios computed from ORLs in several low-metallicity star-forming galaxies compiled by \cite{lopezsanchezetal07, estebanetal14} it has been found that H~{\sc ii} regions in star-forming dwarf galaxies have different chemical evolution histories than the inner discs of spiral galaxies and that the bulk of C in the most metal-poor extragalactic H~{\sc ii} regions should have the same origin as in halo stars (see Figs. 8 and 9 of \cite{estebanetal14}).

\subsubsection{C/O ratios in PNe}

As was pointed out above, there are multiple determinations of C abundance from UV lines in the literature. These determinations, when considered together with detailed analysis of optical spectra covering the same volume of the nebula, provide strong constraints to nucleosynthesis models (see e.~g. \cite{henryetal18}). However, owing to limitations in space, in this text I am going to focus on the determination of C/O ratios from ORLs.

Thanks to the new CCDs with improved efficiency in the blue and the use of large telescopes, several high-quality observations of the C~{\sc ii} $\lambda$4267 RL in PNe have been achieved in the last years (e.~g.  \cite{peimbertetal04, liuetal04, wessonetal05, sharpeeetal07, wangliu07, garciarojasetal12, fangliu11, garciarojasetal18}. 

C/O ratios derived from ORLs combined with other abundance ratios such as N/O or He/H, can set strong constraints to the initial mass of PNe progenitors. This is because different processes occurring at the interior of AGB stars (third dredge-up episodes, hot bottom burning process) activate at different masses and can strongly modify C/O and N/O ratios (see e.~g. \cite{karakaslugaro16} and references therein). These ratios can also provide strong constraints to the physics assumed by nucleosynthesis models, i.~e. the assumption of convective overshooting into the core during the main sequence and the He-burning phases can diminish the mass limit at which He-flashes can occur at a given metallicity (\cite{garciahernandezetal16b}) compared to those models not considering it (\cite{karakaslugaro16}); this has implications for the mass limit at which hot bottom burning can be activated. Additionally, C/O ratios can be used to obtain information about the efficiency of dust formation in C-rich or O-rich environments (see below) and to learn about different dust-formation mechanisms (see \cite{garciahernandezetal16} and \cite{garciarojasetal18}).

\subsection{Abundances of heavy elements in PNe from faint emission lines}
\label{sec:faint}

As I have emphasized several times in this chapter, the increasing efficiency of astronomical detectors as well as the advent of large (8m-10m type) telescopes have boosted the detection of very faint emission lines in photoionized regions. In this section, I will focus on the detection of extremely faint lines in deep spectra of PNe.  

\subsubsection{Faint emission lines of refractory elements}
\label{sec:fe_ni}

The determination of elemental abundances of refractory elements in the spectra of Galactic PNe is not an easy task because the available lines of these elements (mainly iron and nickel) are relatively faint. Additionally, these elements are constituents of dust grains in the ISM and their abundance in photoionized nebulae do not reflect their actual abundances. Several authors have computed abundances of Fe ions in both PNe and H~{\sc ii} regions (\cite{rodriguez96, rodriguez03, delgadoingladaetal09, zhangetal12}) \cite{delgadoingladaetal09} computed detailed Fe abundances for a sample of 28 PNe and found that more than 90\% of Fe atoms are condensed on dust grains. These authors did not find differences between the iron abundances in C-rich and O-rich PNe, suggesting similar depletion efficiencies in both environments. 

\cite{delgadoingladarodriguez14} combined C/O ratios derived from both UV CELs and optical ORLs (as we comment in Sect.~\ref{sec:adf}, they seem to be equivalent) with information obtained from \textit{Spitzer} mid-infrared spectra. They also computed Fe depletions onto dust grains, and found that the highest depletion factors are found in C-rich objects with SiC or the 30 $\mu$m feature in their infrared spectra, while the lowest depletion factors were found for some of the O-rich objects showing silicates in their infrared spectra.

\cite{delgadoingladaetal16} compiled detections of very faint [Ni~{\sc ii}] and [Ni~{\sc iii}] lines in deep spectra of Galactic PNe and H~{\sc ii} regions. They determined the nickel abundance from the [Ni~{\sc iii}] lines using an extensive grid of photoionization models to determine a reliable ionization correction factor (ICF). From the comparison of Fe/Ni ratios with the depletion factor obtained from both [Fe/H] and [Ni/H], they conclude that nickel atoms adhere to dust grains  more efficiently than iron atoms in environments where dust formation or growth is more important.

\subsubsection{Neutron-capture elements}
\label{sec:ncap}

Nebular spectroscopy of neutron(\textit{n})-capture elements (atomic number Z $>$ 30) is a recent field that has seen rapid development in the last 10 years, and holds promise to significantly advance our understanding of AGB \textit{n}-capture nucleosynthesis. Nebular spectroscopy  can reveal unique and complementary information to stellar spectroscopy. 
Observations of PNe provide the first opportunity to study the production of the lightest \textit{n}-capture elements (Z $\le$ 36) and noble gases (Kr and Xe) in one of their sites of origin. Unlike the case of AGB stars, nucleosynthesis and convective dredge-up are complete in PNe, whose envelopes contain material from the last 2$-$3 thermal pulses. 
Accurate computations of \textit{n}-capture elements would shed light on the different scenarios proposed  for the production of these elements and would constrain the chemical yields of low- and intermediate-mass stars for these elements. 

\textit{n}-capture elements were not recognized in any astrophysical nebula until \cite{pequignotbaluteau94} identified emission lines of Br, Kr, Rb, Xe, Ba, and possibly other heavy species in the bright PN NGC\,7027. Since then, a breathtaking number of \textit{n}-capture element emission lines have been identified for the first time in near-infrared (\cite{dinerstein01, sterlingdinerstein08, sterlingetal16, sterlingetal17, madonnaetal18}), UV (\cite{sterlingetal02}) and optical (\cite{sharpeeetal07, garciarojasetal15, madonnaetal17}) spectra of PNe. The new detections have led to a dedicated effort to produce atomic data needed for abundance determinations (e.~g., see \cite{sterlingwitthoeft11, sterlingetal16, madonnaetal18}, and references therein). 
The new photoionization cross-sections and recombination coefficients have been incorporated in photoionization calculations to compute reliable ICFs (e.~g., \cite{sterlingetal15}). The new collisional strengths have been used for abundance determinations of newly detected ions (see \cite{sterlingetal16}, for [Rb~{\sc iv}], \cite{sterlingetal17}, for [Se~{\sc iii}] and [Kr~{\sc vi}], and \cite{madonnaetal18} for [Te~{\sc iii}] and [Br~{\sc v}]). 
Thanks to the fast advances in observations, atomic data determinations and numerical modelling, this field has grown from just 3 PNe with \textit{n}-capture element abundances in 2001 to more than 100 Galactic PNe in 2019 (\cite{sharpeeetal07, sterlingdinerstein08, sterlingetal16, garciarojasetal12, garciarojasetal15, mashburnetal16, sterlingetal17, madonnaetal17, madonnaetal18}).

The importance of deep, high-resolution optical spectrophotometry of PNe to detect faint \textit{n}-capture elements can be understood when comparing the works by \cite{sharpeeetal07} and \cite{garciarojasetal15}. In the first case, several \textit{n}-capture emission lines were discovered in the spectra of 5 PNe, but even at a resolution of $\sim$22000, many features were not unambiguously detected. \cite{garciarojasetal15} took advantage of the very high-resolution (R$\sim$40000) spectrum of NGC\,3918 to clearly identify several ions of Kr, Xe, Rb and Se, testing for the first time the complete set of ICFs for Kr created by \cite{sterlingetal15}. 

The combination of deep optical and near-infrared spectra of PNe has been proved to be a powerful tool to test the predictions of modern AGB nucleosynthesis models as well as to test the accuracy of new atomic data computations and
ICF prescriptions. \cite{madonnaetal17} have combined a deep optical spectrum and a near-infrared spectrum of NGC\,5315, testing for the first time the complete set of ICFs for Se created by \cite{sterlingetal15}. 

Finally, the determination of precise elemental abundances ratios of \textit{n}-capture elements like Kr/Xe or Te/Se from optical and near-infrared spectra are potential indicators of the time integrated neutron flux in the intershell region between the H- and He-burning shells, giving strong constraints to nucleosynthesis models in the thermally pulsing AGB phase \cite{madonna19}.


%
%

\section*{Acknowledgements}

First of all, I want to acknowledge support of the Erasmus+ programme of the European Union under grant number 2017-1-CZ01-KA203-035562. I also acknowledge support from an Advanced Fellowship from the Severo Ochoa excellence program (SEV-2015-0548) and support from the State Research Agency (AEI) of the Spanish Ministry of Science, Innovation and Universities (MCIU) and the European Regional Development Fund (FEDER) under grant AYA2017-83383-P. I am grateful to Dr. Gloria Delgado-Inglada, who kindly provided me the data and the script to produce Fig.~\ref{fig:5} for this chapter. I thank the referee of this chapter, Dr. Roger Wesson, for a detailed report that help to improve the scientific content of the chapter. 

\addcontentsline{toc}{section}{Appendix}

\end{document}